\documentstyle[aps,preprint]{revtex}
\tightenlines

\newcommand{\dpar}[2]{\frac{\partial #1}{\partial #2}}
\def\r{\mbox{\bf r}} 
\def\k{\mbox{\bf k}} 
 
\def\e{{\rm e}}
\def\const{\rm const}
\def\Re{{\rm Re}}
\def\Im{{\rm Im}}
\def\c{c}
\def\b{b}
\def\a{a}
\begin{document}
\input epsf
\draft
\preprint{PURD-TH-99-04, physics/9904067}
\date{April 1999}
\title{Dynamics of lattice spins as a model of arrhythmia}
\author{S. Khlebnikov}
\address{
Department of Physics, Purdue University, West Lafayette, IN 47907, USA
}
\maketitle
\begin{abstract}
We consider evolution of initial disturbances in spatially extended systems 
with autonomous rhythmic activity, such as the heart. We consider
the case when the activity is stable with respect to very smooth
(changing little across the medium) disturbances and construct lattice models 
for description of not-so-smooth 
disturbances, in particular, topological defects; these models are
modifications of the diffusive $XY$ model. We find that when the activity
on each lattice site is very rigid in maintaining its form, the topological 
defects---vortices or spirals---nucleate a transition to a disordered, 
turbulent state.
\end{abstract}
\section{Introduction}
Physical mechanisms underlying many cardiac arrhythmias, in particular
the transition from ventricular tachycardia (VT) to ventricular 
fibrillation (VF), are not fully understood. The ventricular tissue is
known, both experimentally and theoretically, to support long-living
spiral excitations, and it is thought that a breakup of such a spiral
could give rise to a turbulent, chaotic activity commonly associated 
with VF. (Spirals are reviewed in books \cite{books}.) 
A considerable effort is now being directed towards understanding of
these defect-mediated transitions to turbulence within mathematical 
models of ventricular tissue. 
The currently popular approach (reviewed in Ref. 
\cite{Panfilov}) considers a spiral in a patch (or slab) of ventricular tissue; 
the patch is taken in isolation from any pacemaking source. One then 
follows numerically the time evolution of that initial spiral.

In the real beating heart, however, the ventricles are not isolated 
from other regions, and the heart, viewed as a whole, supports 
a (more or less) periodic autonomous activity---the heartbeat itself.
In this case, any defect should  be properly  viewed as a 
disturbance of the normal heartbeat, rather than a structure in isolated 
tissue. In this paper we present some
general results on the evolution of initial disturbances in 
autonomously active media and discuss their possible applications to
cardiac arrhythmias. In particular, we identify a simple mechanism
of defect-induced transition to turbulence in discrete (lattice) 
systems. 
We also find that the more rigid is the system
in maintaining locally the undisturbed form of activity, the more easily 
the transition to turbulence occurs.
This observation can potentially identify a useful therapeutic target.

The assumed lattice structure
need not (though it may) be related to the mechanical structure of the 
medium. The size of the lattice spacing in our models simply represents
the smallest spatial scale on which the rhythmic activity can be 
desynchronized: a region smaller than that scale will necessarily fire
as one. Discrete models of fibrillation have a long history, cf. 
the 1964 model of Moe {\em et al.} \cite{Moe&al}. (Unlike these
authors, though, we do not introduce any frozen inhomogeneity in the
parameters of the medium, apart from the lattice structure itself.)
In addition, the importance of a discrete 
(granular) structure of the medium has been
emphasized in theoretical studies of {\em de}fibrillation 
\cite{defib}.

We introduce an interaction of an excitable region (like the 
ventricles) with a pacemaking region using the following simplified
(not anatomical) model.
We consider a three-dimensional (3d)
slab of simulated medium whose extent in the
$z$ direction  is limited by the planes $z=0$ and $z=L_{z}$. The 
properties of the medium change in the $z$ direction: the region near
$z=0$ is spontaneously oscillatory and represents the pacemaking 
region; the region at larger $z$ is merely excitable and represents
the ventricular tissue. 
The $z$ direction will be also called longitudinal, and the other two
directions, $x$ and $y$, will be called transverse. 
The medium supports a spontaneous 
rhythmic activity, in which an infinite train of pulses propagates 
from small to large $z$. This steady activity is independent of
$x$ and $y$ and is supposed to model the heart's
normal rhythm, in which pulses propagate from the inner surface of the
ventricles out.

The goal of our study was to see what happens if at some instant 
the spontaneous rhythmic activity is disturbed in a spatially 
nonuniform fashion, and then the system is left to itself.
We approach this question in
two steps.  First, we consider the case when the initial disturbance is 
very smooth, i.e. almost uniform across the medium; in particular, 
it captures no topological defects. In this 
case, we expect that {\em locally} the activity rapidly relaxes close
to its undisturbed form. The state can then be described
using a single field $\tau(x,y,z;t)$, which
measures the space- and time-dependent
delay (or advance) in activity among the local regions. 
This field is a phase
variable: it is defined modulo the period $T$ of the 
steady rhythm. For these smooth perturbations, we expect that
the dynamics of $\tau$ at large times
will be universal: it will be described by an
equation whose form (although not the precise values of the coefficients)
does not depend on the details of electrophysiology or on the 
microstructure of the medium. In particular, this large-time
dynamics does not ``see'' the granular structure of the medium.
The form of the equation depends on the symmetries of the medium at
large scales and can be obtained by keeping terms of the lowest 
order in space and time derivatives consistent with the
symmetries. For simplicity, we will assume that at large scales the 
properties of the medium are invariant under translations and rotations
in the $x$--$y$ plane and that $\tau$ does not depend on $z$,
i.e. the disturbance is effectively two-dimensional (2d). (Recall that
$z$ is the direction of propagation of the normal rhythm.)
In this case, the equation describing the large-time dynamics 
has the form
\begin{equation}
  \partial_{t} \theta =
 a \nabla_{2}^{2} \theta + \c (\nabla_{2}\theta)^{2} \; ,
 \label{lwl}
\end{equation}
where the phase $\theta(x,y;t)$ is related to $\tau$ via
\begin{equation}
\theta(x,y;t)=2\pi \tau(x,y;t)/T \; ,
\label{theta0}
\end{equation}
and $a$ and $\c$ are coefficients; $\nabla_{2}$ is the 2d gradient: 
$\nabla_{2}=(\partial_{x}, \partial_{y})$. 

We define a smooth disturbance by the condition
\begin{equation}
|\nabla_2 \theta| \ll 2\pi/L \; ,
\label{smo}
\end{equation}
where $L=\max\{L_x, L_y\}$ is the transverse size of the medium.
Under this condition, the second term in on the right-hand side of
(\ref{lwl}) is much smaller
than the first. We keep it nonetheless, because it is the leading term
that breaks the $\theta\to -\theta$ symmetry. As we will see, 
terms breaking this symmetry play an important role in evolution of
non-smooth disturbances, such as topological defects.
So, it is essential
to establish that the coefficient $\c$ is indeed nonzero.
For smooth disturbances, though, the second term is unimportant,
and eq. (\ref{lwl}) shows that when $a>0$ a smooth initial disturbance
relaxes back to the uniform steady rhythm ($\theta=\const$).
The relaxation process is ordinary diffusion.

It is important to provide a derivation of (\ref{lwl}) from 
an electrophysiological model. In particular, 
that would supply certain values for the yet unknown coefficients
$a$ and $\c$. In Sect. 2 we show how $\theta$ (or $\tau$) can be 
defined within such a model. 
The smaller are gradients of $\theta$, the slower it evolves.
One might think that, given an electrophysiological model,
it should be easy to separate
away the slow dynamics and obtain, quite generally, a closed equation 
for $\theta$. This
task, however, turns out to be far from straightforward, and as of
this writing we have not been able to obtain a general derivation of
(\ref{lwl}); in Sect. 2 we illustrate the nature of the difficulty. 

To establish that the coefficient $\c$ is indeed nonzero, we then have 
resorted to 
the following argument. The simple electrophysiological model that
we consider can be driven, by a choice of the parameters, to a 
critical (bifurcation) point, at which the autonomous rhythmic 
activity is extinguished. Near the critical point, the system can be  
described by a complex Ginzburg-Landau (CGL) model of a complex
order parameter whose phase is our time-delay field $\theta$. For a 
smooth, almost uniform, perturbation, the CGL description reduces to an
equation for $\theta$ alone, and that has the precise form 
(\ref{lwl}), with definite values of $a$ and $\c$. In 
particular, we find that $a>0$ and $\c\neq 0$. As we move away from the 
critical point and towards the form of activity representative of the 
normal heartbeat, the CGL description ceases to be valid. But as it
is difficult to imagine how $\c$ would now suddenly become 
identically zero, we assume that the large-time dynamics 
of $\theta$ is still described by (\ref{lwl}) with a nonzero $\c$. 
We also assume that $a>0$, so that the uniform state is stable. The
electrophysiological model that we use is reviewed in Sect. 3, and the
CGL description is derived in Sect. 4.

The second step of our program is promoting the above 
description of smooth perturbations to a description including not-so-smooth 
perturbations, in particular, topological defects. The latter description 
will not be universal. 
The lack of universality means (by definition) that the description, 
and the type of the resulting dynamics, depend on the microstructure of 
the medium. Because no activity can be fine-grained indefinitely, it is 
natural to assume a granular, or lattice, structure. In Sect. 5, we construct 
lattice models and study their dynamics. In Sect. 6 we summarize our 
results.

\section{Description of smooth disturbances}
In this section we want to show how the slow variable $\theta$, or
equivalently $\tau$, can be defined within the context of an 
electrophysiological model.  This variable evolves arbitrarily slow
in the limit of arbitrarily small gradients; it should not be confused
with ``slow'' recovery variables of electrophysiology.
Our definition of $\tau$ works for any medium
supporting an autonomous periodic activity that is stable with
respect to smooth, almost uniform, perturbations.
For definiteness, we consider here an electrophysiological equation 
of the form
\begin{equation}
\epsilon \ddot{g} - \nabla^{2} \dot{g} - \b \nabla^{2} g - 
F(g,\dot{g};z) = 0 \; .
\label{eqm}
\end{equation}
Overhead dots denote time derivatives, $\nabla$ is the 3d gradient, 
and $\epsilon$ and $\b$ are parameters.
The change in properties of the medium in the $z$ direction is described
by the function $F$, which explicitly depends on $z$.
Eq. (\ref{eqm}) obtains, for instance, when a medium described 
by the two-variable FitzHugh-Nagumo (FHN) model \cite{FHN}
is placed in an external static electric field (we will show that below).
In that case, $g$ is the deviation of the recovery variable of the FHN
model from the static solution.

We consider cases when eq. (\ref{eqm}) (or, more precisely, a suitable
boundary problem based on it) has a periodic in time solution of the form
\begin{equation}
g(\r,t) = \phi(z, t) \; .
\label{phi}
\end{equation}
For example, this solution may
describe a train of pulses propagating in the $z$ direction.
The periodicity means that $\phi(z,t+T)=\phi(z,t)$ for some period $T$.
Notice that, because of the translational invariance of (\ref{eqm})
in time, $\phi(z, t-\tau)$ is also a solution of (\ref{eqm}), for any real 
$\tau$ (albeit with different initial conditions).
We now consider a smooth (in space) perturbation of the periodic
activity described by (\ref{phi}) and assume
that a sufficiently smooth perturbation relaxes back to the periodic state.
After the relaxation has been under way for a while, we expect that 
deviations of $g$ from $\phi$ are already small---except perhaps in the
softest mode, associated with the time translation. We thus seek a 
solution to (\ref{eqm}) of the form
\begin{equation}
 g(\r,t) = \phi(z, t- \tau(\r,t)) + \chi(\r,t) \; ,
\label{sol}
\end{equation}
where $\tau(\r,t)$ is a slowly changing (on the scale of the period 
$T$) function of time: $\dot{\tau} \ll \tau/T$. 
In the limit $\dot{\tau} \to 0$, we should return to 
the solution (\ref{phi}) merely shifted in time, so in this limit 
$\chi$ should vanish. Thus, when $\dot{\tau}$ is small, $\chi$ is also
small, although not necessarily slowly changing.
Because of the periodicity of $\phi$ in time, $\tau(\r,t)$ is a
phase variable: at each spatial point, it is defined modulo the
period $T$. The condition that the perturbation be smooth reduces this
ambiguity to a common shift by $T$ in the entire space.

Note that separation of a perturbation into
$\tau$ and $\chi$ is not completely defined by (\ref{sol}): 
a time-dependent variation in $\tau$
can be absorbed by a variation in $\chi$. This ambiguity can be fixed
by an additional condition---for instance, by requiring that $\chi$ is
orthogonal to $\dot{\phi}$ with respect to a certain inner product.
Eq. (\ref{sol}) together with the additional condition will then 
provide a complete definition of the slow variable $\tau$. 

Now, let us illustrate the nature of the difficulty that arises when 
one tries to derive a closed equation for $\tau$ from eq. (\ref{eqm}).
We substitute (\ref{sol}) into (\ref{eqm}) and expand the right-hand side
to the leading order in small quantities---the function $\chi$ and
the derivatives of $\tau$. The dependence on $\chi$ will be contained 
in an expression of the form ${\hat M}(\phi)\chi$, where ${\hat M}$ is 
a linear operator, which acts on $\chi$ and depends on 
$\phi(z, t- \tau(\r,t))$. Because of the translational invariance of
(\ref{eqm}) in time, the operator ${\hat M}(\phi)$ almost annihilates 
$\dot{\phi}(z, t- \tau(\r,t))$:
\begin{equation}
{\hat M}(\phi)\dot{\phi} \approx 0 \; ;
\label{annih}
\end{equation}
the approximate equality means an equality up to terms of order of the small 
quantity $\partial_{t} \tau$. 
If the operator ${\hat M}(\phi)$ were Hermitean with
respect to an inner product of the form
\begin{equation}
\langle \chi_1 , \chi_2 \rangle = \int_0^{L_z} dz \int_0^T dt w(z,t) 
\chi_1(z,t) \chi_2(z,t) \; ,
\label{inpro}
\end{equation}
for some fixed weight $w(z,t)$, then taking the inner product of (\ref{eqm})
with $\dot{\phi}$ would, to the leading order, project away $\chi$ and produce
a closed equation for $\tau$. In the case of eq. (\ref{eqm}), however, the
explicit form of the operator ${\hat M}$ is
\begin{equation}
{\hat M}(\phi)\chi = \left( \epsilon\partial_t^2 - \nabla^2 \partial_t -
\b \nabla^2 - \dpar{F}{\phi} -\dpar{F}{\dot{\phi}} \partial_t \right) \chi \; ,
\label{M}
\end{equation}
where $F$ is $F(\phi, \dot{\phi}; z)$. This operator is clearly not Hermitean 
with respect to (\ref{inpro}) with $w=1$, and indeed we have not found any 
weight that would render it Hermitean. Thus, we were unable to directly separate
the slow dynamics of $\tau$ from the fast dynamics of $\chi$. While it seems
intuitively clear that the slow dynamics will be described by an equation of the form 
(\ref{lwl}), to establish that the coefficients $a$ and $c$ are indeed both nonzero,
we had to resort to an indirect method, which we describe below.

\section{A model of the heartbeat}
In this section, we describe in some detail the pacemaking mechanism with which
we model the heartbeat. This simple model, based on the two-variable FitzHugh-Nagumo
(FHN) kinetics, will be sufficient for our argument justifying (\ref{eqm})
with nonzero $a$ and $c$.  

Consider a slab of medium described by a FitzHugh-Nagumo model,
\begin{eqnarray}
\epsilon\dpar{E}{t} & = & \nabla^{2} E + f(E) - G \; , \label{eqE} \\
\dpar{G}{t} & = & E - \b G \; , \label{eqG}
\end{eqnarray}
placed in a static uniform external electric field, such as the field of a parallel 
capacitor. Here $E$ is the 
transmembrane voltage, $G$ is the recovery variable, $\epsilon>0$ 
and $\b>0$ are parameters, and $\nabla$ is the 3d gradient. The 
direction of the external field is our longitudinal, or $z$, direction, and the 
slab extends in that direction from $z=0$ to $z=L_z$. The boundary conditions
corresponding to this arrangement are
\begin{equation}
\partial E/\partial z(0) = \partial E/\partial z(L_z) = -{\cal F} \; ,
\label{bc}
\end{equation}
where ${\cal F}$ is a positive constant---the magnitude of the external field.

The boundary problem (\ref{eqE})--(\ref{bc}) has a static solution,
$E_0(z)$, $G_0(z)$.
Deviations from the static solution are $e(\r,t)=E(\r,t)-E_{0}(z)$ and 
$g(\r,t)=G(\r,t)-G_0(z)$.
Excluding the variable
$e$ with the help of (\ref{eqG}), we obtain an equation of the form
(\ref{eqm}) with
\begin{equation}
F(g,\dot{g};z) = f(E_0+ \b g+\dot{g}) - f(E_0) -  g - \epsilon \b\dot{g} \; .
\label{F}
\end{equation}
The explicit dependence of $F$ on $z$ appears through 
the $z$ dependence of $E_0$. 

For a range of ${\cal F}$ the static solution to
(\ref{eqE})--(\ref{bc}) is unstable, for various choices
of $f(E)$, with respect to arbitrarily 
small fluctuations of $E$ and $G$, and the instability gives
rise to an unending time-dependent activity \cite{RK}. 
This will be our pacemaking mechanism.
The corresponding linear stability analysis introduces a number of useful 
definitions, so we briefly go over it here.

Expanding eqs. (\ref{eqE})--(\ref{eqG}) 
to the first order in $e$ and $g$, we obtain
\begin{equation}
\left(\begin{array}{c} \partial e/\partial t \\ 
                                             \partial g/\partial t
           \end{array}
\right)=
\left(\begin{array}{cc} \frac{1}{\epsilon}
     \left( \nabla_{2}^{2} +\dpar{^{2}}{z^{2}} + f'[E_{0}(z)] \right) &
     -{1\over\epsilon} \\
     1 & -\b
           \end{array}
\right)
\left(\begin{array}{c} e \\ g \end{array}
\right) \; .
\label{eqdel}                                           
\end{equation}
This equation should be supplemented by the boundary conditions
\begin{equation}
\dpar{e}{z}(0)=\dpar{e}{z}(L_{z})=0 \; .
\label{bou1}
\end{equation}
Consider eigenfunctions $\psi_{n}(z)$, $n\geq 0$, of the $z$-dependent 
operator in (\ref{eqdel}),
\begin{equation}
\left(-\dpar{^{2}}{z^{2}} -f'[E_{0}(z)] \right) 
\psi_{n}(z)=\lambda_{n}\psi_{n}(z) \; ,
\label{ope}
\end{equation}
with the boundary conditions
\begin{equation}
\dpar{\psi_{n}}{z}(0)=\dpar{\psi_{n}}{z}(L_{z})=0 \; .
\label{bou2}
\end{equation}
We assume that the eigenfunctions $\psi_{n}$ are real and
form a complete orthonormal system on $L_{2}[0, L_{z}]$.

The fields $e$ and $g$ can be expanded in the complete
orthonormal system $\{\psi_{n}\}$:
\begin{eqnarray}
e(\r,t) & = & \sum_{n=0}^{\infty} u_{n}(\r_{2}, t) \psi_{n}(z)  
\label{expe} \; , \\
g(\r,t) & = & \sum_{n=0}^{\infty} v_{n}(\r_{2}, t) \psi_{n}(z)  \; ;
\label{expg}
\end{eqnarray}
here $\r_{2}$ is the two-dimensional coordinate: $\r_{2}=(x,y)$.
Eq. (\ref{eqdel}) then reduces to the following second-order in time
linear equation
\begin{equation}
\ddot{v}_{n}+\left( \b+\frac{\lambda_{n}-\nabla_{2}^{2}}{\epsilon}
\right)  \dot{v}_{n} +
{1\over \epsilon}\left(1+\b[\lambda_{n}-\nabla_{2}^{2}] \right) v_{n}
= 0 \; .
\label{so}
\end{equation}
Eq. (\ref{so}) describes a 
collection of independent oscillators, one for each value of the 
integer $n\geq 0$ and of the 2d wave number $\k$. 
These oscillators have frequencies squared equal to
$\omega_{n}^{2}+ \b k^{2}/\epsilon$
and friction coefficients equal to $\gamma_{n}+k^{2}/\epsilon$, 
where
\begin{eqnarray}
\omega_{n}^{2}& = & (1+\b\lambda_{n})/\epsilon \; ,
\label{ome} \\
\gamma_{n}& = & \b+\lambda_{n}/\epsilon \; .
\label{gam}
\end{eqnarray}
Assuming that the boundary conditions in the $x$--$y$ plane allow
for the $\k=0$ mode, we conclude that
the necessary and sufficient condition for instability is that
\begin{equation}
\lambda_{n}<\max\{-\epsilon \b, -1/\b\}
\label{cond}
\end{equation}
for at least one of the eigenvalues $\lambda_n$. 
This condition corresponds to there being a negative $\omega_{n}^{2}$ 
or a negative $\gamma_{n}$, or both.

The parameter $\epsilon$ sets the ratio of time scales characterizing
changes in the voltage $E$ and in the recovery variable $G$ and is 
typically small. When
$\epsilon <1/b^{2}$, the condition (\ref{cond}) becomes
\begin{equation}
\lambda_{n}<-\epsilon \b \; ,
\label{cond1}
\end{equation}
or equivalently $\gamma_{n} < 0$, where $\gamma_{n}$ is the friction
(\ref{gam}). 

The question that we now address is whether the condition (\ref{cond1}) is
ever satisfied for physiologically relevant values of the parameters.
We choose $\epsilon=0.06$, $\b=0.7$, and $f(E)=6.75E(E-0.25)(1-E)$, 
as recommended in Ref. \cite{Starmer&al} for ventricular tissue with
``normal'' Na and K conductances. The only
other parameter (besides ${\cal F}$) that we need to choose is $L_z$, the 
thickness of the slab in the $z$ direction. This represents the thickness 
of the ventricles in our simplified model. We have done numerical
simulations with $L_z=3.2$.
For lengths, Ref. \cite{Starmer&al} recommends scaling by a factor of 0.5 cm.
A somewhat smaller scaling factor of 0.2 cm is obtained if we equate
the characteristic (``Debye'') length $\xi=0.57$, at which a weak static field gets 
screened inside the medium, to a realistic value of 1 mm. With either scaling,
though, $L_z=3.2$ corresponds to a physical length of order 1 cm. 

To find out if the instability occurs for a given value of ${\cal F}$, one can 
numerically solve the boundary problem (\ref{ope})--(\ref{bou2}) and check
the condition (\ref{cond}). Alternatively, one can numerically
integrate the time-dependent problem
(\ref{eqE})--(\ref{bc}) with initial conditions corresponding to small 
fluctuations near the static solution. This second approach also allows one
to find the form of the time-dependent attractor emerging as the 
instability is cutoff by nonlinear effects,
so we have adopted it. 
For the purposes of this section, it is sufficient to consider initial 
fluctuations that are independent of $x$ and $y$. Using numerical 
integrations of (\ref{eqE})--(\ref{bc}) with such initial conditions 
and with the above values of 
the parameters, we have found that the static
solution is stable as long as ${\cal F} \leq {\cal F}_1\approx 0.4$. 
The value ${\cal F}_{1}$ is the lower critical value, at which the 
static solution first becomes unstable as ${\cal F}$ is increased.
The instability persists as long as ${\cal F}_{1}< {\cal F}<{\cal F}_{2}$
but disappears when ${\cal F}$ reaches the upper
critical value ${\cal F}_{2}\approx 1$. 

The form of the time-dependent attractor, which develops from
small initial fluctuations near the static solution,
is qualitatively different for values of ${\cal F}$ that are close to 
the upper critical field as compared to those elsewhere in the instability
window. These two different forms correspond to propagating versus
nonpropagating activity \cite{RK}. 
In the range ${\cal F}_1 < {\cal F} < {\cal F}_p$,
where ${\cal F}_p$ is somewhat smaller than ${\cal F}_{2}$, the attractor is
an unending train of pulses propagating in the positive $z$ direction. In our
model, this corresponds to the normal heartbeat. On the other hand, when 
${\cal F}_p < {\cal F} < {\cal F}_2$, the development of the instability is 
cut off by nonlinear effects when the deviation from the static solution is too
small to generate a full-fledged pulse. In this case, the entire attractor
lies in the proximity of the static solution. As ${\cal F}$ approaches
${\cal F}_{2}$, the activity is extinguished {\em gradually}: the 
closer is  ${\cal F}$ to ${\cal F}_{2}$, the smaller is the deviation 
from the static solution. This gradual disappearance of activity is 
reminiscent of a second-order phase transition.

\section{The CGL description}
Near the upper critical field, which from now on we will call the
{\em critical point}, the  fields $e(\r,t)=E(\r,t)-E_{0}(z)$ and 
$g(\r,t)=G(\r,t)-G_{0}(z)$ are small
($E_{0}$ and $G_{0}$ denote the static solution).
Expanding the system
(\ref{eqE})--(\ref{eqG}) in $e$ and $g$ so as to retain the leading 
nonlinearities, we obtain
\begin{eqnarray}
\epsilon\dpar{e}{t} &=& \nabla^{2} e + 
f'(E_{0}) e + {1\over 2} f''(E_{0}) e^{2} + {1\over 6} f'''(E_{0})  
e^{3} - g \; , \label{hoe} \\
\dpar{g}{t} &=& e- \b g  \; . \label{hog} 
\end{eqnarray}
As it turns out, the effect of the $e^{2}$ term is relatively 
suppressed and is of the same order as the effect of the
$e^{3}$ term. So, we kept both types of terms in eq. (\ref{hoe}).

Substituting the expansions (\ref{expe})--(\ref{expg}) 
into (\ref{hoe})--(\ref{hog}), we obtain
\begin{eqnarray}
\epsilon\dpar{u_{n}}{t} &=&  (\nabla_{2}^{2}-\lambda_{n}) u_{n} -
 v_{n} - \alpha_{nmm'} u_{m} u_{m'} 
- \beta_{nmm'm''} u_{m} u_{m'} u_{m''} \; , \label{equ} \\
\dpar{v_{n}}{t} & = & u_{n}- \b v_{n} \; ; \label{eqv}
\end{eqnarray}
repeated indices are summed over.
Here $\nabla_{2}$ is the 2d gradient: $\nabla_{2}=(\partial_{x}, 
\partial_{y})$, $\lambda_{n}$ is the eigenvalue
of the Schr\"{o}dinger problem (\ref{ope})--(\ref{bou2}), and $\alpha$ 
and $\beta$ are defined as
\begin{eqnarray}
\alpha_{nmm'} & \equiv & -{1\over 2} 
\int_{0}^{L_{z}} dz f''(E_{0}) \psi_{n} \psi_{m} \psi_{m'} \; ,
\label{alpha} \\
\beta_{nmm'm''}& \equiv & -{1\over 6} 
\int_{0}^{L_{z}} dz f'''(E_{0}) \psi_{n} \psi_{m} \psi_{m'} 
\psi_{m''} \; . \label{beta}
\end{eqnarray}

We stay closely enough to the critical point, so that on that side of 
it where the static solution is unstable there will be only 
one $\lambda_{n}$ satisfying the instability condition (\ref{cond}).
That will be $\lambda_{0}$. In what follows we only consider cases 
when $\epsilon < 1/\b^{2}$. Then, the instability condition takes the form
\begin{equation}
\gamma_{0} < 0 \; ,
\label{cond2}
\end{equation}
where $\gamma_{0}= \b + \lambda_{0}/\epsilon$ is the friction
coefficient (\ref{gam}) for $n=0$.
The closer the system is to the critical point, the smaller is 
$|\gamma_{0}|$. We make it small
enough, so that the frequency squared (\ref{ome}) with $n=0$ (and
hence with all $n>0$ as well) is positive and much larger than
$\gamma_{0}^{2}$:
\begin{equation}
\omega_{0}^{2} = 1/\epsilon - b^{2}  + b\gamma_{0} 
\gg \gamma_{0}^{2} \; .
\label{ome1}
\end{equation}
The large positive $\omega_{0}$ sets the time scale of rapid
oscillations of $u_{n}$ and $v_{n}$.

We now want to show that when the system is sufficiently close
to the critical point its dynamics on time scales of order of and larger than
$|\gamma_{0}|^{-1}$ is described by a 2d complex
Ginzburg-Landau (CGL) model. The field $\Psi(\r_{2},t)$ of this CGL 
model is defined via the expansion
\begin{equation}
v_{0}(\r_{2},t) = \left( {\Psi \over \b-i\omega_{0}}\e^{-i\omega_{0}t} 
+{A_{0} \Psi^{2}\over \b-2i\omega_{0}} \e^{-2i\omega_{0}t} + {\rm c.c.} 
\right) + {C_{0} \over \b} \Psi^{\dagger} \Psi
+ \ldots \; ,
\label{psi}
\end{equation}
where the omitted terms are higher harmonics, proportional to the third 
and higher powers
of $\exp(\pm i\omega_{0}t)$; c.c. means complex 
conjugate. The coefficients  $A_{0}$ and $C_{0}$ are in principle series in
$\Psi^{\dagger} \Psi$, but near the critical point $\Psi$ is small, 
and to the leading order $A_{0}$ and $C_{0}$ can be 
regarded as constants, which will be determined later.
The definition (\ref{psi}) separates away the rapid 
oscillations with frequency $\omega_{0}$ and its multiples
and, in this sense, is 
analogous to a transition to the nonrelativistic limit in field theory.

The CGL description is obtained by substituting (\ref{psi}) into eqs.
(\ref{equ})--(\ref{eqv}), expanding to the third order in $\Psi$, and finally 
retaining only terms that 
contain $\exp(\pm i\omega_{0}t)$ in powers 0, 1, and 2. One can verify
that terms omitted in (\ref{psi}) will not contribute to the resulting 
equation.
For instance, terms proportional to $\exp(\pm 3i\omega_{0}t)$ are of
order $\Psi^{3}$; to convert them into terms of lower order in
$\exp(\pm i\omega_{0}t)$ one will need to multiply
them by at least one power of $\Psi$ or $\Psi^{\dagger}$, which will
make them of the fourth order in $\Psi$.

The CGL description allows us to consider disturbances of the
uniform activity that satisfy the conditions
\begin{equation}
\dot{\Psi} = O(\Psi^{3}) \; , ~~~~~\nabla_{2}^{2}\Psi = O(\Psi^{3}) \; .
\label{smo1}
\end{equation}
These are less restrictive than the smoothness condition (\ref{smo}),
which now takes the form
\begin{equation}
|\nabla_{2} \Psi|/|\Psi| \ll 2\pi/L \; .
\label{smo2}
\end{equation}
In particular, unlike (\ref{smo1}), the condition (\ref{smo2}) 
explicitly prohibits topological defects, which are centered at
zeroes of $|\Psi|$.
Under the more restrictive condition (\ref{smo2}), the CGL dynamics 
reduces, at sufficiently large times, to dynamics 
of the phase of $\Psi$ alone.

To the third order in $\Psi$,
$u_{0}$ is obtained from (\ref{eqv}) and (\ref{psi}) as
\begin{equation}
u_{0}(\r_{2},t) = C_{0} \Psi^{\dagger} \Psi + 
\left( \Psi\e^{-i\omega_{0}t}+ A_{0} \Psi^{2}\e^{-2i\omega_{0}t} 
+ {\dot{\Psi} \over \b-i\omega_{0} } \e^{-i\omega_{0}t} + {\rm c.c.} 
\right) + \ldots \; ,
\label{u0}
\end{equation}
where dots again denote higher harmonics.
As will be checked a 
posteriori, $v_{n}$ and $u_{n}$ with $n>0$ are of order $\Psi^{2}$.

In this approximation,
eqs. (\ref{equ})--(\ref{eqv}) with $n=0$ become
\begin{eqnarray}
\epsilon\dpar{u_{0}}{t} &=&  (\nabla_{2}^{2}-\lambda_{0}) u_{0} -
v_{0} - \alpha_{000} u_{0}^{2} - 2\alpha_{00\nu} u_{0} u_{\nu}
- \beta_{0000} u_{0}^{3} \; , \label{equ0} \\
\dpar{v_{0}}{t} & = & u_{0}- \b v_{0} \; ,\label{eqv0}
\end{eqnarray}
where $\nu>0$, while for $n=\nu>0$ they become
\begin{eqnarray}
\epsilon\dpar{u_{\nu}}{t} &=&  -\lambda_{\nu} u_{\nu} -
v_{\nu} - \alpha_{\nu 00} u_{0}^{2} \; , \label{equnu} \\
\dpar{v_{\nu}}{t} & = & u_{\nu}- \b v_{\nu} \; . \label{eqvnu}
\end{eqnarray}
We see that in this approximation the modes
with $n=\nu>0$ are damped linear oscillators driven by the external 
force proportional to $u_{0}^{2}$. For the purpose of calculating
$u_{\nu}$, it is sufficient to take $u_{0}^{2}$ computed to the 
second order in $\Psi$:
\begin{equation}
u_{0}^{2} = 2 \Psi^{\dagger}\Psi 
+\left( \Psi^{2} \e^{-2i\omega_{0} t} +{\rm c.c.} \right) 
+ O(\Psi^{3}) \; .
\label{force}
\end{equation}
Then, the solution for $u_{\nu}$ at large times  is 
\begin{equation}
u_{\nu} = A_{\nu}\Psi^{2}\e^{-2i\omega_{0} t} 
+ A_{\nu}^{*} (\Psi^{\dagger})^{2}\e^{2i\omega_{0} t} 
+ C_{\nu} \Psi^{\dagger} \Psi + O(\Psi^{3}) \; ,
\label{unu}
\end{equation}
where
\begin{eqnarray}
A_{\nu} & = & - \alpha_{\nu 00} 
\left(\lambda_{\nu} - 2i\epsilon \omega_{0} +
{1\over \b -2i\omega_{0}} \right)^{-1}  \; , \label{A} 
\\
C_{\nu} & = & - 2 \alpha_{\nu 00} 
\left(\lambda_{\nu} + 1/\b \right)^{-1} \; . \label{C}
\end{eqnarray}
Substituting this expression for $u_{\nu}$ into eq. (\ref{equ0}) for
$u_{0}$ we see that the only effect of the modes with $n>0$ is  a local
(in space and time) renormalization of the dynamics of the $n=0$ mode.

To complete our derivation of the CGL description, we now turn to eq. 
(\ref{equ0}) and compose separate equations for different  powers of 
$\exp(-i\omega_{0}t)$. The equations for the zeroth and second powers 
give expressions for $C_{0}$ and $A_{0}$ that are of the same form
as (\ref{A})--(\ref{C}) but with $\nu$ everywhere replaced by 0.
The equation for the first power then gives the CGL equation
\begin{equation}
\dot{\Psi} = D\nabla_{2}^{2} \Psi - {1\over 2} \gamma_{0} \Psi - 
s \Psi^{2} \Psi^{\dagger} \; ,
\label{cgl}
\end{equation}
where the complex diffusion coefficient is
\begin{equation}
D = {1\over 2\epsilon} \left( 1 + {i\b\over \omega_{0}} \right) \; ,
\label{D}
\end{equation}
and the complex coupling constant is
\begin{equation}
s = D \left(
-2 \sum_{n=0}^{\infty} \alpha_{00n}^{2} \left(
{2\b\over \epsilon \omega_{n}^{2}}
+\left(\lambda_{n} - 2i\epsilon \omega_{0} +
{1\over \b -2i\omega_{0}} \right)^{-1}  \right)
+ 3\beta_{0000}
\right) \; .
\label{s}
\end{equation}
Recall that the condition of instability of the 
static solution is $\gamma_{0} < 0$, and near the critical point 
$|\gamma_{0}|$ is small.

Spatially uniform activity near the critical point (for 
$\gamma_{0}<0$) is described by the  following solution of (\ref{cgl}):
\begin{equation}
\Psi_{0}(t) = \rho_{0} \exp(-i s_{I} \rho_{0}^{2} t) \; ,
\label{psi0}
\end{equation}
where $\rho_{0} = (|\gamma_{0}|/2s_{R})^{1/2}$; $s_{R}$ and $s_{I}$
are the real and imaginary parts of $s$. Of course,
this solution exists only when $s_{R}> 0$. 
For a smooth perturbation of this uniform activity (which, in particular, 
contains no topological defects), we can define the modulus 
$\rho(\r_{2}, t)$ and the phase $\theta(\r_{2},t)$ via
\begin{equation}
\Psi(\r_{2},t) = \rho(\r_{2}, t) \exp(-i s_{I} \rho_{0}^{2} t + 
\theta(\r_{2}, t) ) \; .
\label{theta}
\end{equation}
Substituting this into eq. (\ref{psi}) shows that $\theta$ measures the 
phase shifts in periodic activity among local regions, so it is
precisely the variable that we defined in Sect. 2. As
the modulus $\rho$ relaxes close to $\rho\approx \rho_{0}$ everywhere 
in the 2d space, eq. (\ref{cgl}) reduces to 
an equation for the phase $\theta$ alone. That equation is of 
the form (\ref{lwl}), with $a=\Re D$, and $\c=-\Im D$.

\section{Construction of lattice models}
As we move away from the critical point and towards the form of  
activity that is more representative of the normal heartbeat, 
the CGL description ceases to be valid. 
Nevertheless, we expect that eq. (\ref{lwl}) will still 
apply for sufficiently smooth perturbations. That is because $\theta$ is the 
only variable that can change arbitrarily slowly (for arbitrarily small
gradients), and the two terms on the right-hand side of (\ref{lwl}) are
the only two terms of the lowest (second) order in gradients that are
consistent with the symmetries of our model and the assumption that
$\theta$ does not depend on $z$. Moreover, we now have a reason to 
believe that both coefficients $a$ and $\c$ will be nonzero: 
we have seen that they were both nonzero near the critical point, and it is 
hard to imagine how either of them would vanish identically when we move
away. So, we consider eq. (\ref{lwl}) to be reasonably well justified.

The next step is to build upon (\ref{lwl}) to construct models 
that would apply to not-so-smooth perturbations of the normal
rhythm, in particular, to those containing topological defects. 
As we consider perturbations of progressively smaller spatial scales, 
there are
two effects that lead to deviations from (\ref{lwl}). On the 
one hand, the granular (lattice)  structure of the medium becomes 
important; on the other hand, the local form of activity deviates
from its unperturbed form, so that other variables besides $\theta$ 
come into play. We have found that the resulting dynamics depends
crucially on which of these two effects becomes important first, i.e.
at larger spatial scales. In what follows, we 
contrast the corresponding two types of the dynamics.
Finding out which one is realized in a specific medium will 
require a detailed electrophysiological model. The required model
will have to include the details
of the granular structure, so it cannot be a simple continuum model
of the type we used to justify eq. (\ref{lwl}).

First, consider the case when the local activity is very {\em 
rigid} in maintaining its form. That means that each 
grain---or lattice site---still 
carries on essentially the undisturbed activity, so the field $\theta$ 
remains the only requisite variable. In this case, the dynamics is
described by a model of classical lattice $XY$ spins. For definiteness,
we consider here a model on a square lattice, with interactions 
restricted to the nearest  neighbors (NN). (Similar results were
obtained for a model that includes interactions of next-to-nearest neighbors.)
We take the model equation in the form
\begin{equation}
\partial_{t} \theta_{i} = h^{-2} \sum_{j\in {\rm NN}(i)} \left[
\a \sin(\theta_{j} - \theta_{i}) 
+ \c (1 - \cos(\theta_{j} - \theta_{i})) \right] \; .
\label{eqlat}
\end{equation}
The index $i$ labels the sites of a 2d square lattice, and
$h$ is the lattice spacing. Matching to the long-wave limit
(\ref{lwl}) identifies $\a$ and $\c$ in (\ref{eqlat}) with 
those in (\ref{lwl}). 

Near the critical point, $\c/\a= - \b/\omega_{0}$,
which is proportional to the small $\sqrt{\epsilon}$. Away from the 
critical point, however, there is no reason to expect 
$|\c/\a|$ to be small, and we need to explore the dynamics 
of the model for diverse values of this ratio. We assume that $\a>0$ 
and set $\a=1$  by a rescaling of time.

When $\c=0$, eq. (\ref{eqlat}) becomes the
usual diffusive $XY$ model. This model has stable topological 
defects---vortices and antivortices. A nonzero $\c$ gives these
defects a rotation (clockwise or counterclockwise, depending on the
sign of $\c$), so vortices and antivortices become spirals. By
numerically integrating (\ref{eqlat}), we have
found that for small values of $|\c|$ these spirals are stable---or 
at least no instability could be detected during finite times of our 
computer runs. 

\begin{figure}
\leavevmode\epsfysize=3.0in \epsfbox{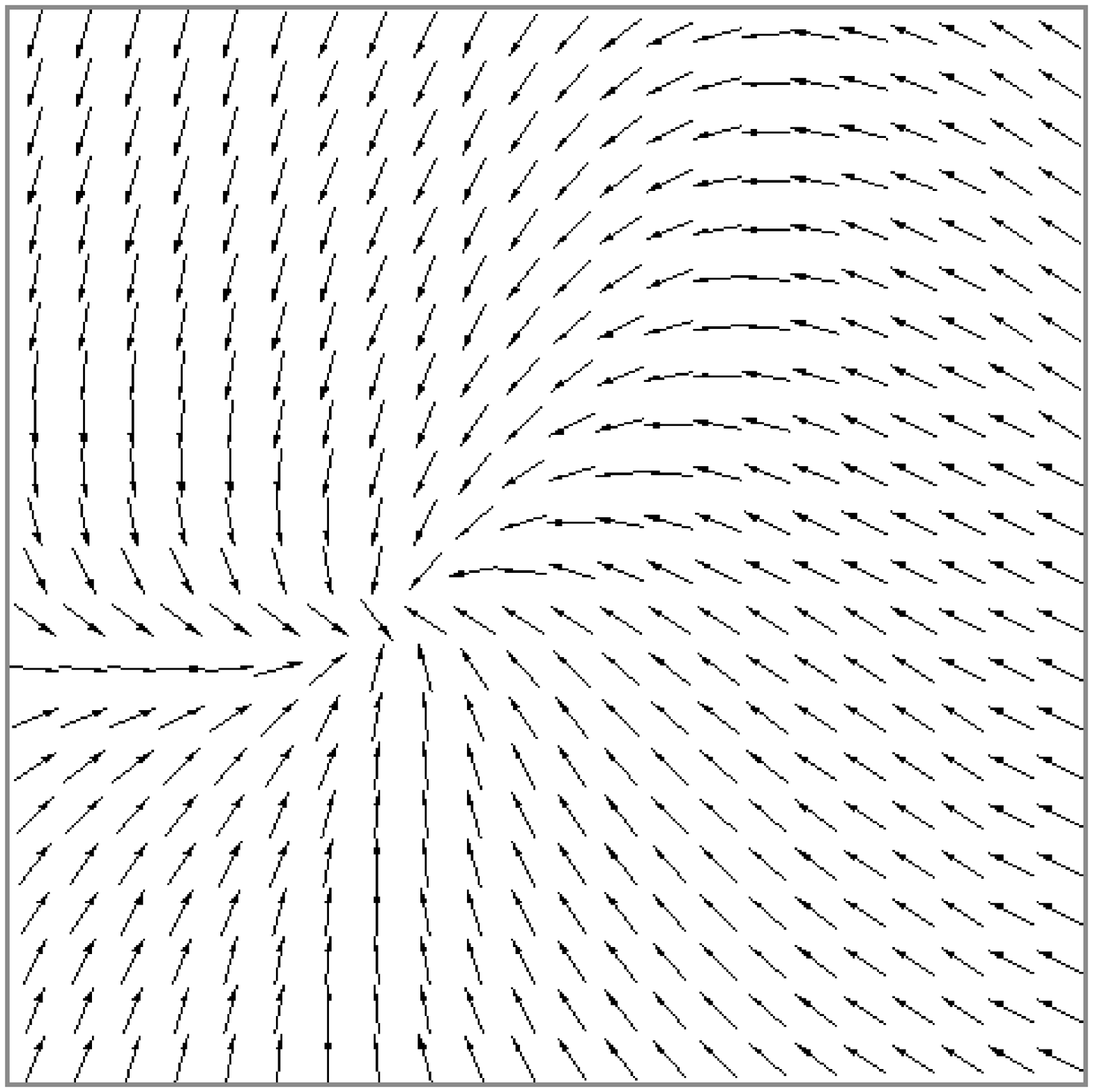}
\caption{Field distribution at $t=0$.}
\label{fig:init}
\end{figure}

\begin{figure}
\leavevmode\epsfysize=3.0in \epsfbox{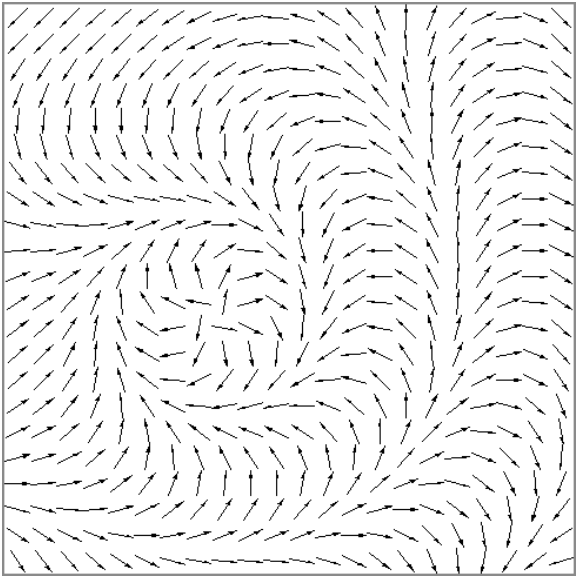}
\caption{Field distribution at $t=20$ in the model
(\ref{eqlat}) with $\a=1$ and $\c=-0.5$.}
\label{fig:spiral}
\end{figure}

As $|\c|$ is increased, the spirals become more tightly wound
and at a sufficiently large $|\c|$ they become unstable.
Formation of a tightly wound but still stable spiral is illustrated
by Figs. \ref{fig:init}, \ref{fig:spiral}. 
Fig. \ref{fig:init} shows an 
initial state, containing a single vortex,
and Fig. \ref{fig:spiral} shows the spiral that develops from that initial
state for $\a=1$ and $\c=-0.5$.
The values of $\theta$ at a given time are represented as 
directions of lattice spins, as measured clockwise from 12 noon \cite{dl}.
These results were obtained via Euler's explicit time-stepping scheme
on a $33\times 33$ lattice with side length $L=10$
and discretized Neumann boundary conditions. For picture clarity, only a
$22\times 22$ square is shown.

\begin{figure}
\leavevmode\epsfysize=3.0in \epsfbox{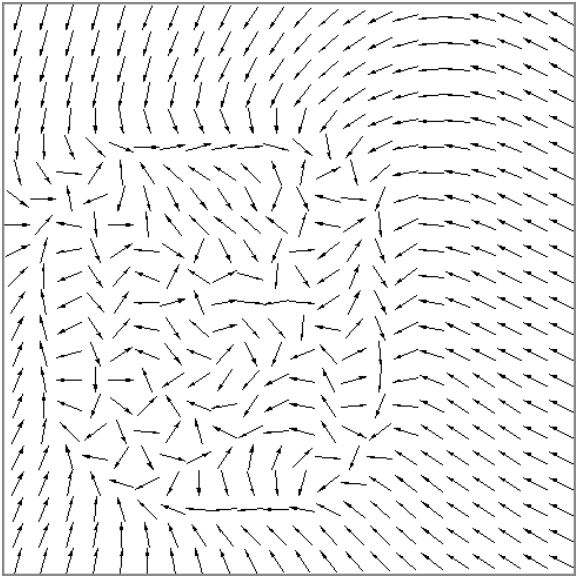}
\caption{Field distribution at $t=0.3$ in the model
(\ref{eqlat}) with $\a=1$ and $\c=-2$.}
\label{fig:bubble}
\end{figure}

Evolution of an unstable defect is illustrated 
by Fig. \ref{fig:bubble}. This picture was obtained for 
$\a=1$ and $\c=-2$ on the same lattice and with the same
initial condition as Fig. \ref{fig:spiral}.
The center of the defect now serves as a nuclei of a new phase,
a featureless turbulent state. A bubble of the new phase originates
at the center of the defect and rapidly grows, eating up the ``normal'' phase, 
until the new phase occupies the entire volume.
As far as we can tell, the resulting turbulent state is 
persistent. Fig. \ref{fig:bubble}
shows the bubble during its growth. This growth is indeed so rapid that
the initial vortex
does not have time to fully develop into a spiral, although some fragments 
of spiral structure can be seen near the wall of the bubble.
A patch of the turbulent state is seen inside the bubble, away from the wall.
When the turbulent state occupies the entire volume, it remains disordered:
directions of the spins are uncorrelated beyond a few lattice spacings.
In addition, spins in the turbulent state
rapidly change their directions with time.

Next, we consider a case when the local activity is {\em flexible}, 
i.e. it readily changes its form in response to a short-scale 
perturbation. For instance, we can supply the lattice spins with a 
variable length by making 
$\theta$ the phase of a complex field $\Phi=|\Phi| \exp(i\theta)$.
This introduces an additional degree of freedom associated with
$|\Phi|$. As an illustration, consider $\Phi$ that obeys a complex
Ginzburg-Landau (CGL) equation:
\begin{equation}
\dpar{\Phi}{t} = D\nabla^{2}\Phi + r \Phi (1- |\Phi|^{2} ) \; ,
\label{cgl1}
\end{equation}
where $D=\a- i\c$; for simplicity we take the coupling $r$ 
to be real: $r> 0$.
We can now discretize eq. (\ref{cgl1}) on a 2d square lattice of spacing 
$h$ and vary the parameter $r$ in relation to 
$h^{-2}$. At large $r$, the modulus $|\Phi|$ freezes out
at $|\Phi| \approx 1$, and we obtain a lattice model of $\theta$ alone,
in the spirit (although not necessarily of the exact form) of eq. (\ref{eqlat}).
At small $r$, the natural size of a defect's core will be set by
$(|D|/r)^{1/2}$, rather than the lattice spacing, so we expect that
the discretization will be irrelevant, and the dynamics 
will approach that of the continuum 2d CGL model.
This latter model has 
spiral solutions that are at least core-stable in a certain range of its 
parameters \cite{cgl}.
Numerically integrating discretized eq. (\ref{cgl1}), we have found
that by varying $r$, for a fixed
$\c/\a$, one can interpolate between the 
unstable spirals of a lattice model with fixed-length
spins and the stable spirals of the continuum CGL model.

\section{Conclusion}
In this paper we tried to implement consistently the idea that 
a disturbance in the normal heartbeat can be viewed as 
a collection of ``clocks'',  each of which measures
the local phase of the activity.
In conjunction with the view that the heart has a granular (or 
lattice) structure, this idea leads to a description of the heart via
lattice models of classical spins. Our main results are as 
follows. 

(i) Assuming that sufficiently smooth (almost uniform across the 
medium)
disturbances of the normal rhythm relax back to it, one can write down
a universal description of this relaxation process. 
Universality means that the form of the equation is independent of
details of microscopics.
For a simplified model of the heartbeat, and disturbances depending 
only on the transverse (with respect to the direction of pulse propagation) 
coordinates, the universal description is eq. (\ref{lwl}). Although we 
have not derived this equation in the general case, we have justified it 
by presenting a derivation near a critical (bifurcation) point.

(ii) For not-so-smooth disturbances, including topological 
defects, dynamics begins to depend on the assumed lattice structure 
and the details of electrophysiology. In particular, we have found that 
it depends strongly on how rigid the local activity is in maintaining 
its form. When the activity is very rigid (fixed length spins), the 
system, for a range of the parameter space, is prone to a 
defect-induced instability, which leads to a disordered, 
turbulent state. 

We expect that the local rigidity of the medium (in the above sense)
will depend on 
its longitudinal size (the thickness of the ventricles) and on
the electrophysiological parameters, such as Na and K conductances.
Since, according to our results, the local rigidity plays such an 
important role in the transition to turbulence (fibrillation), its 
dependence on the parameters may serve to identify useful therapeutic 
targets.

\end{document}